\documentclass[twocolumn,runningheads]{svjour2}
\smartqed  
\usepackage{graphicx}
\newcommand\apj{{ApJ}}
\newcommand\apjs{{ApJS}}
\newcommand\aap{{A\&A}}
\newcommand\ssr{{Space~Sci.~Rev.}}

\journalname{Astrophysics and Space Science}
\begin{document}

\title{Demystifying an unidentified EGRET source by VHE gamma-ray observations}



\author{Olaf Reimer        \and
        Stefan Funk       
}

\authorrunning{Olaf Reimer \& Stefan Funk} 

\institute{Olaf Reimer \at
           W.W. Hansen Experimental Physics Laboratory \& \\
           Kavli Institute for Particle Astrophysics and Cosmology\\
		   Stanford University \\
           Stanford, CA 94305-4085, USA\\
           Tel.: +1-650-724-6819, Fax: +1-650-725-2463\\
  \email{olr@stanford.edu}  
           \and
           Stefan Funk \at
           Kavli Institute for Particle Astrophysics and Cosmology, \\
           SLAC \\
		   Menlo Park, CA 94025, USA\\
           Tel.: +1-650-926-8979, Fax: +1-650-926-5566\\
  \email{Stefan.Funk@slac.stanford.edu} 
}

\date{Received: date / Accepted: date}

\maketitle

\begin{abstract}
In a novel approach in observational high-energy gamma-ray astronomy, observations carried 
out by imaging atmospheric Cherenkov telescopes provide necessary templates to pinpoint the 
nature of intriguing, yet unidentified EGRET gamma-ray sources. Using GeV-photons detected by CGRO 
EGRET and taking advantage of high spatial resolution images from H.E.S.S.\ observations, we 
were able to shed new light on the EGRET observed gamma-ray emission in the Kookaburra complex, 
whose previous coverage in the literature is somewhat contradictory. 3EG\,J1420--6038 very likely 
accounts for two GeV gamma-ray sources (E\,$>$\,1 GeV), both in positional coincidence with the 
recently reported pulsar wind nebulae (PWN) by HESS in the Kookaburra/Rabbit complex. 
PWN associations at VHE energies, supported by accumulating evidence from observations in the 
radio and X-ray band, are indicative for the PSR/plerionic origin of spatially coincident, but 
still unidentified Galactic gamma-ray sources from EGRET. This not only supports the already 
suggested connection between variable, but unidentified low-latitude gamma-ray sources with 
pulsar wind nebulae (3EG\,J1420--6038 has been suggested as PWN candidate previoulsy), it also 
documents the ability of resolving apparently confused EGRET sources by connecting the GeV emission 
as measured from a large-aperture space-based gamma-ray instrument with narrow field-of-view 
but superior spatial resolution observations by ground-based atmospheric Cherenkov telescopes, 
a very promising identification technique for achieving convincing individual source identifications 
in the era of GLAST-LAT.
 
\keywords{EGRET \and Data Analaysis \and GLAST \and Simulations \and Pulsars \and Pulsar Wind Nebulae }

\PACS{98.70.Rz, 97.60.Gb, 95.85.Pw, 98.70.Rz}
\end{abstract}

\section{The EGRET detected gamma-ray emission in the Kookaburra complex}

The EGRET instrument aboard Compton Gamma-Ray Observatory initially reported high-energy gamma-ray 
emission at E\,$>$\,100 MeV in the First EGRET catalog as GRO\,J1416--61 \cite{Fic94}, thereby 
confirming a positional coincidence with the previously detected COS-B source 2CG311--01 \cite{Swa81}. 
This COS-B source was already suspected and investigated as potential PSR candidate \cite{Ami83}. 
With the accumulating data and the improved understanding of the instrument response during the EGRET mission, 
the gamma-ray source was refined on basis of a 2-year exposure, and labeled 2EG\,J1412--6211 \cite{Tho95}. 
On the basis of additional and privileged on-axis exposure in the third year of the 
EGRET operations, a new source 2EGS\,J1418--6049 was reported \cite{Tho96}, 
a 7 sigma detection at E\,$>$\,100 MeV at the location (l=313.31, b=0.29). A catalog compiled from EGRET detected
photons with energies $>$1 GeV \cite{Lam97} lists a 6 sigma source GEV\,J1417--6100 at the location 
(l=313.18, b=0.14), "identified" with the unidentified EGRET source 2EGS\,J1418--6049. 
A similar GeV-study \cite{Rei97} found the gamma-ray excess located at 
(l=313.49, b=0.38). These reports were superseded with the appearance of the results from 
the Third EGRET catalog \cite{Har99}, which reports two sources at E $>$ 100 MeV 
in the vicinity: 3EG\,J1410--6147 at (l=312.18, b=-0.35), tentatively associated with the 
previously seen source 2EG\,J1412--6211, and 3EG~J1420-6038, a 6.5 sigma detection at 
(l=313.63, b=0.37), tentatively associated with the previously seen source 2EGS\,J1418--6049. 
Both sources were dubbed "C", meaning source confusion may affect flux, significance, or 
position of the accordingly flagged catalog sources. This source was also listed as 
coinciding with the E\,$>$\,1 GeV source GEV\,J1417--6100.

\begin{figure}
\centering
  \includegraphics[width=0.48\textwidth]{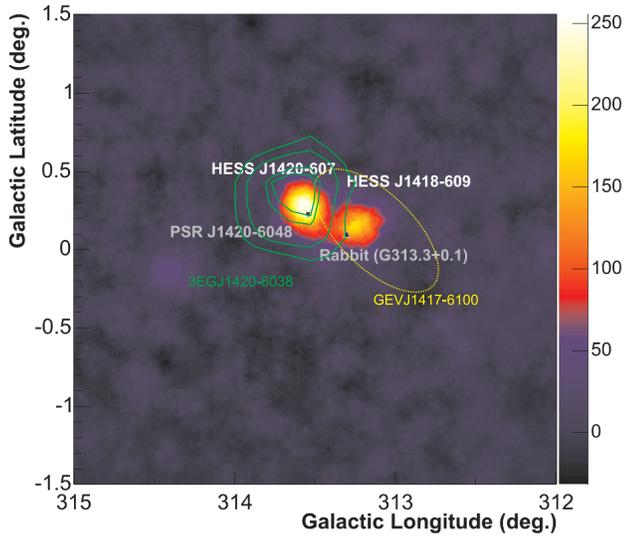}
\caption{The Kookaburra complex as seen in high-energy gamma-rays between 100 MeV and $\sim$25 TeV. Overlaid on the 
smoothed excess map from H.E.S.S.\ observations \cite{Aha06} are the source location confidence contours for 
3EG\,J1420--6038 \cite{Har99}, and GEV\,J1417--6100 \cite{Lam97}. The discrepancy is obvious, in particular since both EGRET 
source locations share photons in the E $>$ 1 GeV regime due to the size of the EGRET point spread function. 
Taken both published results at its face value, it would indicate that the flux ratio at the location 
of the H.E.S.S.\ source location changes dramatically around $\sim$1 GeV, or the GeV emission shifts 
its emission peak. Although interesting hypothesis on its own, we investigate here the consistency between 
the 3EG catalog result and the earlier GEV catalog analysis.}
\label{fig:1}       
\end{figure}

These EGRET detections subsequently received considerable attention. Although the spatial 
coincidence with a Supernova remnant was already noticed earlier \cite{Stu95}. An reassessment \cite{Cas99},
of the suggested association between 2EGS\,J1418--6049 and SNR\,312.4--0.4 was made, concluding that this source 
is transient in nature and its variability makes it unlikely to be associated with the Supernova remnant or 
isolated pulsar, thus putting it among the candidates for a new class of yet unidentified Galactic sources. 
At the same time, the region was investigated under the hypothesis of its PSR/PWN nature \cite{Rob99}. 
VLA and hard-X-ray observations were used to study the multifrequency properties 
of sources in the region, and since then, the region was dubbed "Kookaburra" to account for the 
very distinctive synchrotron emission features seen at 20 cm. The "Kookaburra" region was found 
to contain two "wings" of non-thermal emission, whos most prominent features were later referred to as 
K3 and "Rabbit".

ASCA data taken at the location of the GeV source were used to study the pulsar PSR\,J1420--6048 as its putative
counterpart \cite{Rob01}. Although PSR\,J1420--6048 ranks high among the energetic 
pulsars in $\rm {\dot E /d^2}$, the 68 ms periodicity could not be established in the EGRET detected 
gamma-ray photons yet. Shortly before the H.E.S.S.\ observations of the region were announced, 
a double pulsar wind nature, corresponding to K3 and the Rabbit, of the nonthermal emission was 
suggested \cite{Ng05} from newly obtained Chandra and XMM observations. That has proven 
to be the most plausible counterpart hypothesis, since H.E.S.S.\ observations impressively 
confirmed the nonthermal nature in the Kookaburra complex of extended radio and X-ray 
sources, which both have the characteristics of PWN \cite{Aha06}. The confirmation of a PWN hypothesis 
by VHE gamma-ray astronomy through the detection of HESS\,J1420--607 and HESS\,J1418--609 accounts nicely 
for another intriguing problem concerning unidentified EGRET sources: There was not a single firm 
identification achieved among the population of variable, presumably Galactic unidentified gamma-ray sources. 
If variability is used as a discriminator to distinguish between SNR/PSR and PWN, we have at least 
three candidate source populations to account for variable gamma-ray emission from unidentified 
gamma-ray sources at locations close to the Galactic equator:\\
(1) Active Galactic Nuclei shining through the Plane, e.g. 3EG\,J2016+3657 \cite{Muk00},\\
(2) PWN, suggested both from studying the Crab off-pulse emission \cite{Jag94}, as well as from numerous 
positional coincidences between energetic pulsars and unidentified gamma-ray sources, and \\ 
(3) Microquasars, as impressively confirmed by the detection of LS5039 with H.E.S.S.\ \cite{Aha05} (possible 
associated with 3EG\,J1824--1514) and LSI 61$^\circ$303 with MAGIC 
\cite{Alb06} (possible associated with 3EG\,J0241+6103). 
Since the H.E.S.S.\ observations we are in possession of precise nonthermal emission templates for the 
Kookaburra complex, and one can attempt to solve the obvious discrepancy between the EGRET source 
locations as reported in the 3EG catalog and the GEV catalog, respectively.


\section{Why re-analysing the EGRET data?}

\begin{figure}
\centering
  \includegraphics[width=0.48\textwidth]{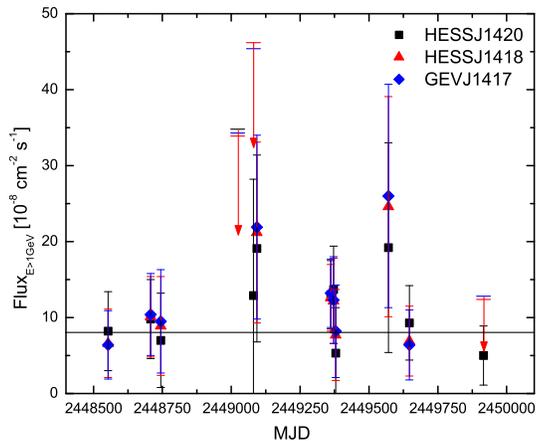}
\caption{Fluxes determined on three test positions in the EGRET E\,$>$\,1 GeV data from the Kookaburra/Rabbit 
complex. Variability is apparently not as pronounced as reported from E\,$>$\,100 MeV analysis in the region.}
\label{fig:2}       
\end{figure}

\begin{figure}
\centering
  \includegraphics[width=0.50\textwidth]{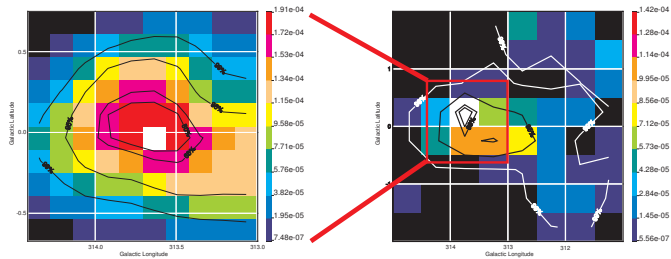}
\caption{Likelihood test statistics map of the region of the Kookaburra complex between 2 and 4 GeV. The source 
location is at (l=313.71, b=0.04), close to the previously reported location of 3EG\,J1420--607 at (l=313.63, b=0.37), 
but not confirming the location at GEV\,J1417--6100 at (l=313.18, b=0.14). The EGRET source indeed coincides 
with HESS\,J1420--607. There is substantial excess emission towards HESS\,J1418--609, though.}
\label{fig:3}       
\end{figure}

Fig.1 shows the smoothed excess map from the Kookaburra complex as seen by H.E.S.S.\
PSR\,J1420--6048 and the Rabbit (G313.3+0.1) are marked. Overlaid are the 
50, 68, 95, and 99\% confidence level for the maximum likelihood location of 3EG\,J1420--6038 
\cite{Har99}, and the 95\% containment error ellipse of the maximum likelihood 
location of GEV\,J1417--6100 \cite{Lam97}. Apparently, it remains unclear whether: 
(a) 3EG\,J1420--6038 and GEV\,J1417--6100 are one and the same source or not. Given the size 
of the EGRET psf at E\,$>$\,100 MeV and E\,$>$\,1 GeV, they certainly share photons; 
(b) the location of 3EG\,J1420--6038 coincides preferably with HESS\,J1420--607; 
(c) the location of GEV\,J1417--6100 coincides preferably with HESS\,J1418--609. 
The 3EG catalog generally gives information on sources detected above an analysis threshold of 
E\,$>$\,100 MeV. However, during catalog compilation analysis results in the 300-1000 MeV and E\,$>$\,1 GeV energy 
bands were also considered, therefore the discrepancy remains to be disentangled. We remark, 
that in the 3EG catalog it is explicitly written that the likelihood test statistic maps in 
different energy bands "were compared, and the one which produced the smallest error contours was chosen 
to represent the source position, as long as the significance was greater than 4 sigma, 
a level chosen to reflect a substantial degree of confidence in the detection." This was 
exactly the case for 3EG\,J1420--6048, whose location originated from an E\,$>$\,1 GeV analysis. 
Therefore we don't even have a discrepancy between EGRET analysis results obtained at different
analysis thresholds (E\,$>$\,100 MeV, and E\,$>$\,1 GeV, respectively), but in fact directly between 
two independently determined source locations from E\,$>$\,1 GeV photons!\\
We therefore analyzed EGRET viewing periods throughout the CGRO mission, where the 
Kookaburra complex was within 25$^\circ$ on-axis, which we could consistently analyzed with 
the EGRET narrow field-of-view point spread function. Furthermore, we tested several 
selection criteria in order to get a hint if some of the viewing periods are indicative 
for underlying systematic problems, like an extreme correction factor in the spark chamber 
efficiency normalization \cite{Esp99}. The analysis was performed with 
EGRET data from viewing periods vp0120, 0230, 0270, 2080, 2170, 2180, 3140, 3150, 
3160, 4020, 4025, 4240, in various energy bands (E\,$>$\,100 MeV, E\,$>$\,300 MeV, E\,$>$\,1 GeV, 
E\,$>$\,4 GeV, 1 GeV\,$<$\,E\,$<$\,4 GeV, 2 GeV\,$<$\,E\,$<$\,4 GeV, 2 GeV\,$<$\,E\,$<$\,10 GeV etc. We 
find the best compromise between improved instrument psf and better discrimination of a hard 
spectrum source against the Galactic diffuse emission towards higher energies, and sufficiently 
large number of photons within the sample to be analyzed in the energy range 2 GeV\,$<$\,E\,$<$\,4 GeV, 
which we report here. Furthermore, we used also fixed test positions at the location of the newly 
detected H.E.S.S.\ sources to check what an EGRET source modeled at an alternative location than determined 
by an unbiased max likelihood may provide to an understanding of the situation at GeV energies. 
Ultimately, we will draw a flux ratio for the gamma-ray emission at E\,$>$\,1 GeV at the position of the 
two H.E.S.S.\ sources in order to refine the spectral energy distribution for improved multiwavelength modeling.


\section{Results of the EGRET re-analysis}

Fig.2 shows the lightcurve of the analyzed EGRET data for E\,$>$\,1 GeV. At GeV energies, 
the source does not exhibits the strong variability as previously reported for E\,$>$\,100 MeV 
\cite{McL96}, \cite{Tor01}, \cite{Nol03}. This can be due to our privileged selection of viewing 
periods towards optimal observing conditions, but also intrinsic to photons measured at E\,$>$\,1 GeV 
like changes among the different emission components/processes in this complex region if the source 
is of composite nature. Fig.3 gives the EGRET maximum likelihood analysis result of the region, 
overlaid with the 50, 68, 95, and 99\% source location confidence contours. We conclude that the 
source location of GEV\,J1417--6100 was imprecisely determined, and there is a consistent picture 
achieved where a dominant source in the region indeed coincides with the reported location in the 
3EG catalog. This source is confused with a less intense gamma-ray source towards the location of the 
Rabbit, which itself is below a conservative detection threshold to be reported individually as detection 
on the basis of the EGRET data. The maximum likelihood position of the GeV emission spatially coincides 
with HESS\,J1420--607. At approximately or less then 1/3 of the GeV flux coincident with HESS\,J1420--607, there 
is excess emission consistent with the location of HESS\,J1418--609.

\begin{figure}
\centering
  \includegraphics[width=0.48\textwidth]{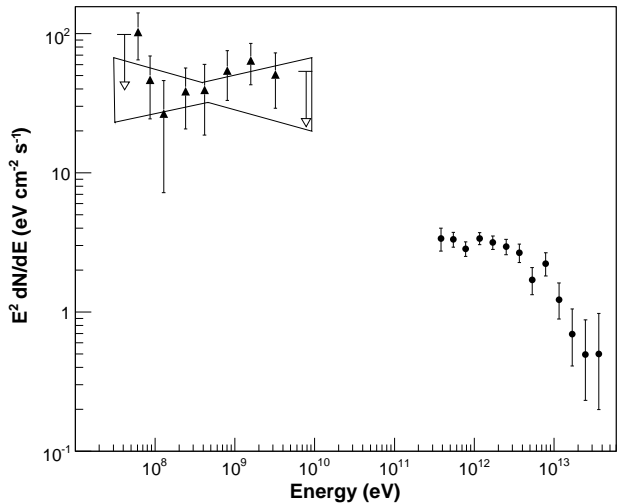}
\caption{EGRET and H.E.S.S.\ spectra measured in the Kookaburra complex. Note that the H.E.S.S.\
data only contain the contribution from both HESS\,J1420--607, and HESS\,J1418--609, whereas the EGRET data contain 
the GeV photons of the whole region according to EGRETs larger instrumental point-spread-function.}
\label{fig:4a}       
\end{figure}


\section{Expectations for GLAST-LAT}

With these results at hand, we aim to predict how the Kookaburra complex might be seen 
by the Gamma-Ray Large Area Space Telescope (GLAST). Assuming that there is a connection 
between the H.E.S.S.\ source and the GeV emission, one can model the spectral energy distribution 
(SED) in terms of a leptonic acceleration scenario in which gamma-rays are produced by 
Inverse Compton scattering of high-energy electrons on background photons. The parameters 
for this model are constrained by the H.E.S.S.\ spectral points, the ASCA X-ray data on the 
PWNe and the total EGRET flux for 3EG\,J1420--6048. Figure 4 shows the high energy part 
of the SED with the EGRET and the H.E.S.S.\ observations. Is it assumed that the EGRET spectrum 
contains an unknown combination of the flux from the pulsar PSR\,J1420--6048 and the flux 
from the PWNe. In a first attempt to estimate the signal seen by GLAST, 100\% of the total 
flux of the region has been assigned to the two H.E.S.S.\ pulsar wind nebulae, and a simulation 
of 5 years of LAT observation (including the diffuse background) has been performed. Using 
the H.E.S.S.\ 2-D map as shown in Fig.1 as a template for the location of the photons and 
leptonic emission scenarios as shown in Fig.5 as a template for the energy distribution of the 
photons, a simulation of the region in a PWN scenario was obtained. The resulting spectral points 
for such GLAST observations are shown in Figure 5. We note, that very likely the contribution 
from the pulsar will be distinguishable through its periodicity, given that 3EG\,J1420--6038 is 
already a strong GeV emitter as reported from EGRET observations. Therefore, the shown PWN 
scenario resembles an OFF-pulse analysis which will be feasible with GLAST-LAT if periodicity 
from a pulsar could be established. The pulsed emission component may dominate the already 
measured GeV emission entirely.

\begin{figure}
\centering
  \includegraphics[width=0.48\textwidth]{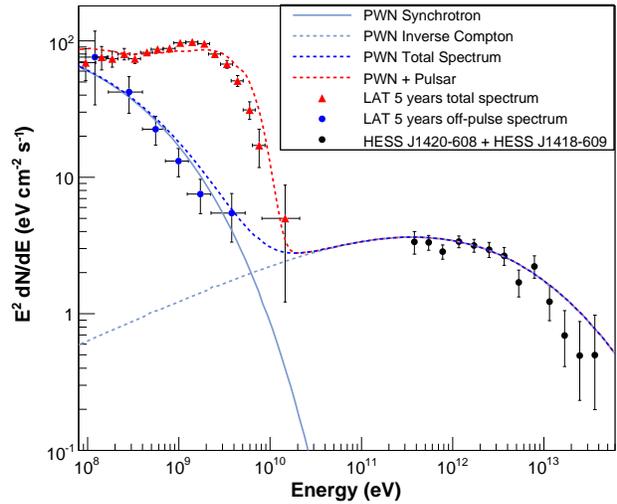}
\caption{A model of the SED of the whole region in terms of a one-zone leptonic emission model. 
The triangles show the expected signal for a 5-year GLAST orbit. Please note that due to a 
non-optimized analysis technique, this spectrum should be treated as a conservative estimate 
of what is to be expected from GLAST.}
\label{fig:4b}       
\end{figure}



\end{document}